\documentclass[amsmath,amssymb,onecolumn, showkeys, a4paper,14pt]{revtex4}
\usepackage{amsmath}
\usepackage{amssymb}
\usepackage{amsmath,amsthm,amssymb,amscd}
\usepackage{latexsym}
\usepackage{indentfirst}
\usepackage{subfigure}
\usepackage{graphicx}
\usepackage{extarrows}
\usepackage{bm}
\usepackage{multirow}
\usepackage[colorlinks,linkcolor=blue,hyperindex,dvipdfm]{hyperref}

\date{\today}

\begin{document}

\title{
  How does a protein reach its binding locus: sliding along DNA chain or not?
}

\author{Jingwei Li, Yunxin Zhang} 
\affiliation{Shanghai Key Laboratory for Contemporary Applied Mathematics, Centre for Computational Systems Biology, School of Mathematical Sciences, Fudan University, Shanghai 200433, China.}

\begin{abstract}
In gene expression, various kinds of proteins (such as polymerase or transcription factor) need to bind to specific locus of DNA. Although sophisticated experiments have been done according to this process, it is still not clear how these proteins find their target locus. Are these target-search processes completed mainly by 3-dimensional diffusion in cell space or with the aid of 1-dimensional sliding along DNA chain?  Previous studies have shown that sliding along DNA chain may help to increase the search efficiency. While recent experiments also found that the length of DNA sequence has little influence on the search time. In this study, the mean first-passage time (FPT) of protein binding to its target locus on DNA chain is discussed by a chain-space coupled model. In which the cell space is simply represented by a 2-dimensional rectangular lattice and the DNA chain is simplified to a 1-dimensional lattice with length $L$. Our results show that the mean FPT has power law relation with the 2-dimensional diffusion constant approximately. The 1-dimensional diffusion constant has a critical value, with which the mean time spent by a protein to find its target locus is almost independent of the binding rate of protein to DNA chain and the detachment rate from DNA chain. Which implies that, the frequency of protein binding to DNA and the sliding time on DNA chain have little influence on the search efficiency, and therefore whether or not the 1-dimensional sliding on DNA chain increases the search efficiency depends on the 1-dimensional diffusion constant of the protein on DNA chain. This study also finds that only protein bindings to DNA loci which are close to the target locus help to increase the search efficiency, while bindings to those loci which are far from the target locus might delay the target binding process. As expected, the mean FPT increases with the distance between the initial position of protein in cell space and its target locus on DNA chain. While our results show that the mean FPT does not change monotonically with the distance between the initial position of protein and the DNA chain. To know how a protein reaches its target locus, i.e., binding the target through its adjacent loci of DNA or directly binding through its nearest neighbor position in the cell space, the direct binding probability, which can be regarded as one index to describe if the 1-dimensional sliding along DNA chain is helpful to increase the search efficiency is calculated. Our results show that the influence of 1-dimensional sliding along DNA chain on the search process depends on both diffusion constants of protein in cell space and on the 1-dimensional DNA chain.
\end{abstract}


\maketitle

\section{Introduction}
During gene expression, specific protein molecules, such as RNA polymerase and transcription factor, need to recognize and bind to certain loci on DNA chain, which usually lie in promoter domain \cite{AlbertsJohnson2007,Buc1985RNApromoter,Dehaseth1998,Jensen1998,Sanchez2011,VanniniCramer2012,SaeckerRecorddeHaseth2011}. These binding processes are important for biological systems to regulate gene expressions \cite{Berg1987Regulatory,Alper2005LibraryReuse,Cox2007Reuse,Rhodius2010OtherPredict,LeeMinchinBusby2012}. The mechanism of how a protein reaches its target locus on DNA is a basic biophysical problem, and has been extensively studied both experimentally and theoretically \cite{ElfLiXie2007,TafviziHuang2011,HalfordMarko2004,KolomeiskyPhysicsof2011}. Nevertheless, the mechanism of this target search process remains unclear
\cite{MirnySlutsky2009,KolomeiskyPhysicsof2011}.
In references, various methods of theoretical analysis have been presented to try to explain this fast search process in cells \cite{RoeBurgess1984,FriedmanGelles2012,BujardRodriguezChamberlin1982}, which is usually called facilitated diffusion (FD) due to its high efficiency. Including the approach of lowering dimensionality \cite{BergWinter1981,WinterHippel1981,HalfordMarko2004,MirnySlutsky2009}, electrostatic effects \cite{HalfordAnend2009}, correlations between 3D and 1D motions \cite{CherstvyKolomeiskyProtein2008,ZhouRapid2011,KolomeiskyPhysicsof2011}, transitions between different chemical states \cite{ReingruberHolcman2011,TafviziHuang2011}, as well as bending
fluctuations and hydrodynamics \cite{HansenNetz2010}.

In this study, we mainly want to show that if the one-dimensional sliding of protein along DNA chain attributes to this search process for a target on DNA. Or in other words, if the one-dimensional sliding is essential to increase the search efficiency, and can shorten the search time effectively. In previous studies \cite{MirnySlutsky2009,Kabata1993Science,Park1982JOBC,Park1982JOBC2,Singer1987JOBC,Singer1988JOBC,Guthold1999BJ,Harada1999BJ,Ricchetti1988PNAS,SakataSogawa2004PNAS,BaiSantangeloWang2006,WangGreeneSingle2011,BergWinter1981}, the protein molecule is thought to first bind to a nonspecific locus of DNA chain through three-dimensional diffusion in cell space, and then slide along the one-dimensional DNA chain to reach the target locus (or binding site). Corresponding theoretical analysis showed that, with the help of this one-dimensional sliding along DNA chain (lowering dimensionality), the target search efficiency of a protein can be greatly increased \cite{BergWinter1981,WinterHippel1981,HalfordMarko2004,MirnySlutsky2009}. So it is believed that the one-dimensional sliding along DNA chain is essential to accelerate this search process, and consequently is important to gene expression in cell. However, observations in recent experiments \cite{HalfordAnend2009,Friedman2013PNAS,Wang2012NSMB} showed that most proteins reached their target loci without long-range one-dimensional sliding along the DNA chain, and therefore this target search process is mainly completed through three-dimensional diffusion in cell space. So, the mechanism of how proteins in cells, including polymerase and transcription factor, can find their binding sites on DNA chain effectively remains unclear.

One can image that the search mode of proteins depends on both cell environment and protein properties, especially the diffusion constants in cell space and along DNA chain, as well as the binding/unbinding rates of proteins to/from the DNA chain. For example, with low values of one-dimensional diffusion constant along DNA chain and high values of unbinding rate, proteins will reach their target loci mainly through diffusion in cell space, and vice versa. Thus the contribution of one-dimensional sliding along DNA chain to this target search process of proteins is influenced by both the cell environment and protein properties.
\begin{figure}
  \centering
  \includegraphics[width=12cm]{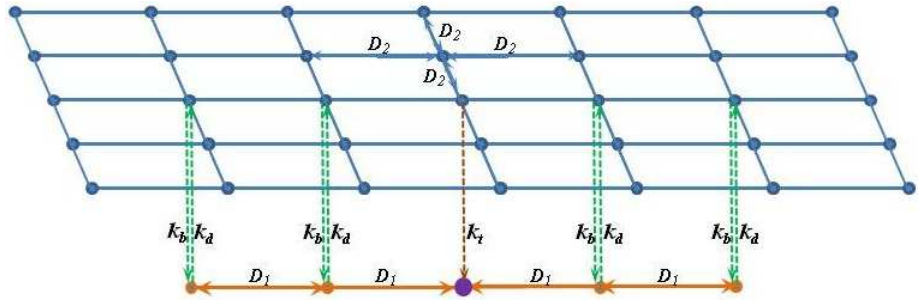}\\
  \caption{Diagram of the chain-space coupled model used in this study. The DNA chain is simplified as a one-dimensional lattice with length $L$ ($L=5$ in this diagram), and assumed to lie at the middle of the cell space. The cell space is simplified as a rectangular two-dimensional lattice with size $M\times N$ ($M=7,N=5$ here). The target locus of a protein is assumed to lie at the center of the DNA chain, i.e. the lattice site $(L+1)/2$ for an odd number $L$. A protein molecule in cell space can walk randomly between adjacent lattice sites with rate $D_2$, or along the DNA chain with rate $D_1$. Proteins can bind to or detach from the DNA chain randomly with rate $k_b$ and $k_d$, respectively. The binding rate of a protein molecule from cell space to the target locus on DNA chain is denoted by $k_t$.}\label{modeldiagram2}
\end{figure}

To show how the search mode of a protein molecule is influenced by diffusion constants and binding/unbinding rates, a similar model as the one used in \cite{Kolomeisky2012TJOCP} is employed in this study. In which the DNA chain is simplified to a one-dimensional lattice with length $L$, with each DNA locus represented by a lattice site. The cell space is simplified to a two-dimensional rectangular lattice with  $M\times N$ lattice sites, see Fig. \ref{modeldiagram2}. Five parameters are included in our model, one-dimensional diffusion constant along DNA chain $D_1$, two-dimensional diffusion constant in cell space $D_2$, binding/unbinding rate of protein molecule to/from nonspecific binding site $k_b$ and $k_d$, and direct binding rate from cell space to the target locus on DNA chain $k_t$. If the diffusion constant $D_2$ is extremely large compared with binding rates $k_t$ and $k_b$, the mean first-passage time (FPT) of a protein molecule from cell space to target binding site can be obtained explicitly. For general cases, numerical computations are employed.

Our results show that there is a critical value of the diffusion constant $D_1$, with which the mean FPT is insensitive to values of binding/unbinding rates $k_b$ and $k_d$, unless $k_b$ and $1/k_d$ are extremely large. In other words, with this critical value, how often and how long a protein molecule slides along the DNA chain have almost no influence on the target search efficiency. For values of diffusion constant $D_1$ which is larger than this critical value, the one-dimensional sliding along DNA chain will be helpful to increase the search efficiency, and vice versa. Which means that for large values of $D_1$, the search efficiency will increase with the binding rate $k_b$ while decrease with the unbinding rate $k_d$. While for small values of $D_1$, search efficiency will decrease with rate $k_b$ but increase with rate $k_d$.
Meanwhile, in this study, the influences of two-dimensional diffusion constant $D_2$, the length $L$ of DNA chain, and the distance $d_{\rm target}$ ($d_{\rm chain}$) between the initial position of a protein molecule and its target locus (DNA chain) are also discussed. Finally, a method to calculate the probability $P^{\rm direct}$ that a protein molecule reaches the target DNA locus directly, i.e. bind to the target locus through the adjacent positions in cell space but not from its nearest neighboring loci on DNA chain, is also presented. Here, $P^{\rm direct}$ can be regarded as one index to evaluate the importance of the one-dimensional sliding along DNA chain in the target search process of proteins.

This paper is organized as follows. The model used in this study will be briefly introduced in Section \ref{model}, and then theoretical methods to get the mean FPT for large limit values of diffusion constant $D_2$ will be presented in Section \ref{secTheoForLarK2}. For general cases, results obtained by numerical computations will be given in Section \ref{numericalFPT}. In Section \ref{diffusiondependence}, the dependence of direct binding probability $P^{\rm direct}$ on diffusion constants $D_1$ and $D_2$ will be discussed theoretically. Finally, concluding and remarks will be presented in the last section. 

\section{Model description}\label{model}
In our model, the DNA chain is simplified to a one-dimensional lattice with length $L$, with each DNA locus represented by a lattice site. While the cell space is simplified to a two-dimensional rectangular lattice with size $M\times N$, see Fig. \ref{modeldiagram2} for a schematic depiction in which $L=5$ and $M=7,N=5$. The diffusion of protein molecule is then simplified to random walks between adjacent lattice sites. For convenience, this study assumes that the target DNA locus lies at the center of the DNA chain, i.e. the lattice site ${(L+1)}/{2}$ of DNA chain. Meanwhile, the DNA chain is assumed to be horizontal and lies at the center of the rectangular lattice (in this study, $L$, $M$, $N$ are always chosen to be odd numbers).

To reach its target binding site on DNA chain, the protein molecule first walks randomly on the two-dimensional rectangular lattice with rate $D_2$. When it reaches lattice sites adjacent to the DNA chain, it may either bind to the nearest DNA site with rate $k_b$ if this site is not the target locus, or bind with rate $k_t$ if the nearest site is the target locus, or just walks away randomly with rate $D_2$ in the rectangular lattice. The observations in recent experiments \cite{Friedman2013PNAS} showed that the binding rate of RNA polymerase to promotor is usually larger than that to other regions of DNA chain, therefore $k_t$ is generally larger than $k_b$. After binding to DNA chain, the protein molecule will walk randomly along the one-dimensional lattice with jumping rate $D_1$. At sites which are not the target DNA locus, the protein may detach into the cell space (i.e. the rectangular lattice) again with rate $k_d$.

In this study, periodic boundary conditions are used for random walk in the two-dimensional rectangular lattice, which means that the left and bottom boundaries of the rectangular lattice are connected with the right and top boundaries, respectively. In all the following numerical calculations, $M=N=101$ are used, and each site of the rectangular lattice is denoted by its position coordinate $(i,j)$, with $i$ the column index and $j$ the row index of rectangular lattice, see Fig. \ref{modeldiagram2}.

\section{Mean first-passage time for large limit values of rate $D_2$}
\label{secTheoForLarK2}
In this section, we will derive the expression of mean FPT for the special cases in which the two-dimensional diffusion rate $D_2$ is large enough. For convenience, we define several notations as follows. Let $U_i$ be the mean FPT of a protein which initiates from site $i$ of DNA chain to reach the target site $({L+1})/{2}$. Let $Q_i$ be the splitting probability that a protein initiated at DNA site $i$ reaches the target site without unbinding from the DNA chain, and $T_i$ be the conditional mean FPT of $Q_i$. Let $P_{ij}$ be the splitting probability that a protein initiated at DNA site $i$ detaches from the DNA site $j$ without reaching the target site. It is obvious that $Q_i+\sum_{j\ne(L+1)/2} P_{ij}=1$. Let $S_i$ be the conditional mean FPT of $1-Q_i$. Actually, $S_i$ is the conditional mean FPT that a protein initiated at DNA site
$i$ detaches from the chain (at any site) without reaching the target site. Let $R_i$ be the mean FPT of a protein from position $(({M-L})/{2}+i,({N+1})/{2})$ (corresponding to the position of DNA site $i$) in the 2D rectangular lattice to DNA chain. Let $O_{ij}$ be the splitting probability that a protein initiated at position $(({M-L})/{2}+i, ({N+1})/{2})$ binds to the DNA chain at site $j$. For definitions of (conditional) mean FPT, splitting probability, see \cite{Redner2001CUP}.

With the above notations, one can show that
\begin{eqnarray}\label{UieqScale}
U_i=Q_iT_i+(\sum_{j=1}^LP_{ij})S_i+\sum_{j=1}^LP_{ij}R_j+\sum_{k=1}^L\sum_{j=1}^LP_{ij}O_{jk}U_k,
\end{eqnarray}
where $1\le i\le L$.
Or it can be written as the following matrix form,
\begin{eqnarray}
U=Q\circ T +(Pe)\circ S+PR+POU.
\end{eqnarray}
Here, $U$, $Q$, $T$, $S$, $R$ are $L\times 1$ vectors with components $U_i$, $Q_i$, $T_i$, $S_i$, $R_i$ respectively. $P$ and $O$ are $L\times L$ matrices with components $P_{ij}$ and $O_{ij}$. All components of the $L\times 1$ vector $e$ equal one. $Q\circ T$ means an $L\times 1$ vector with components $Q_iT_i$. $(Pe)\circ S$ is defined in the same way.

Let $F_i$ be the mean FPT of a protein from position $(({M-L})/{2}+i, ({N+1})/{2})$ in cell space to the target site $(L+1)/2$ of DNA chain. Then we have
\begin{eqnarray}
F_i=R_i+\sum_{j=1}^LO_{ij}U_j.
\end{eqnarray}
Or in matrix form $F=R+OU$, with $F$ an $L\times 1$ vector with components $F_i$.

Given $Q$, $T$, $P$, $S$, $R$ and $O$, the mean FPT $U$ can be obtained from Eq. (\ref{UieqScale}). But explicit expressions of $O$ and $R$ are difficult to obtain. Meanwhile, even if $O$ and $R$ are obtained, one still need to solve the inverse of an $L\times L$ matrix to get $U$. Therefore, in this section we only discuss the limit cases in which the two-dimensional diffusion rate $D_2$ is much larger than binding rates $k_b $ and $k_t$ of protein to DNA chain, i.e. $D_2\gg k_b,k_t$. General cases will be discussed in the next section by numerical computations. For these limit cases, one can easily show that $R\approx eR_0$ with $R_0=(\bar{k}_bL/MN)^{-1}={CMN}/{(Lk_b)}$, $O_{ij}=C/L$ for $j\neq{(L+1)}/{2}$, and $O_{ij}=Ck_t/(Lk_b)$ for $j={(L+1)}/{2}$. Where $\bar{k}_b=[(L-1)k_b+k_t]/L$ is the average binding rate of a protein to DNA chain, and $C=k_b/\bar{k}_b$. Note that $U_{{(L+1)}/{2}}=0$, thereby
\begin{eqnarray}
OU=(C/L)EU=C\bar{U}e,
\end{eqnarray}
where $E$ is an $L\times L$ matrix with all elements equal to one, and $\bar{U}={(\sum_{i=1}^L U_i)}/{L}$. From definitions one can easily show that $Pe=e-Q$, so
\begin{eqnarray}
U=Q\circ T +(Pe)\circ S+(e-Q)R_0+C(e-Q)\bar{U},
\end{eqnarray}
and
\begin{eqnarray}\label{largek2F1}
F=(R_0+C\bar{U})e.
\end{eqnarray}
It can be proved that (see Section \ref{appendixTheoretical} of the Supplementary Material)
\begin{eqnarray}\label{heQmean}
Q\circ T +(Pe)\circ S=\frac{1}{h}(e-Q).
\end{eqnarray}
Thus,
\begin{eqnarray}\label{expressionU}
U=\frac{1}{h}(e-Q)+(e-Q)R_0+C(e-Q)\bar{U}.
\end{eqnarray}
Define $H:=e^T(e-Q)/L=\overline{e-Q}$, where $e^T$ is the transpose of $e$, i.e. is a $1\times L$ vector with all components equal to one. In fact, $H$ is the average value of all components of vector $e-Q$, and therefore also denoted by $\overline{e-Q}$. Multiplying both sides of Eq. (\ref{expressionU}) by ${e^T}/{L}$, we obtain
\begin{eqnarray}
\bar{U}=\frac{H}{h}+R_0H+CH\bar{U},
\end{eqnarray}
so
\begin{eqnarray}\label{expressionbarU}
\bar{U}=\frac{\frac{H}{h}+R_0H}{1-CH}.
\end{eqnarray}
Substituting Eq. (\ref{expressionbarU}) into Eq. (\ref{largek2F1}), we get
\begin{eqnarray}\label{expressionFatlargek2final}
F=(\frac{R_0+\frac{C}{h}H}{1-CH})e.
\end{eqnarray}
For the calculations of $Q$ and $H$, see Section \ref{appendixTheoretical} of the Supplementary Material.
\begin{figure}
  \centering
  \includegraphics[width=10cm]{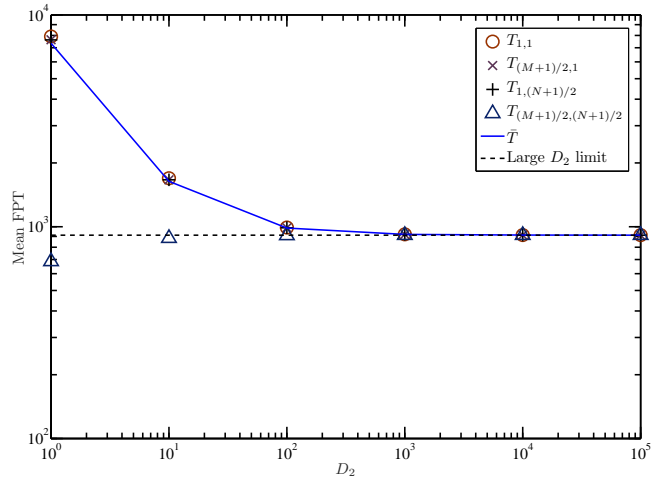}\\
  \caption{The average of mean FPT $\bar{T}$ (see Eq. (\ref{averageMFPT}) for the definition of $\bar{T}$), and four typical examples of mean FPT $T_{i,j}$, with $(i,j)=(1,1), ((M+1)/2,1), (1,(N+1)/2), ((M+1)/2,(N+1)/2)$, respectively, as functions of two-dimensional diffusion constant $D_2$. In calculations, $M=N=101$, $L=51$, $k_b=1$, $k_d=1$, $k_t=10$, and $D_1=1$ are used. One can find that all numerical values of $\bar{T}$ and $T_{i,j}$ tend to the theoretical one as $D_2\to\infty$, see discussions in Section \ref{secTheoForLarK2}.
  }\label{FatLargeD_220160328}
\end{figure}

Let $T_{i,j}$ be the mean FPT of a protein from position $(i,j)$ in cell space to the target binding site $(L+1)/2$ on DNA chain. Define
\begin{eqnarray}\label{averageMFPT}
\bar{T}=\frac{\sum_{i=1}^M\sum_{j=1}^N T_{i,j}}{MN},
\end{eqnarray}
as the average of mean FPTs over the 2D space with prior uniform distribution. In order to validate Eq. (\ref{expressionFatlargek2final}), examples of $T_{1,1}$, $T_{{(M+1)}/{2},1}$, $T_{1,{(N+1)}/{2}}$, $T_{{(M+1)}/{2}, {(N+1)}/{2}}$, and $\bar{T}$ are numerically calculated and plotted in Fig. \ref{FatLargeD_220160328}. The results show that they all tend to the theoretical value given by Eq. (\ref{expressionFatlargek2final}) as rate $D_2\to\infty$.

\section{Mean first-passage time to reach target binding site: general cases}\label{numericalFPT}
In previous section, the mean FPT of a protein molecule to find its target binding site in DNA chain has been obtained explicitly for large value limit of diffusion rate $D_2$. In the following, we will discuss the general cases but through numerical computations, see Section \ref{CalmFPTSPcmFPT} of the Supplementary Material for some details of the numerical method used in this study. We will mainly focus on the influences of diffusion rates $D_1$ and $D_2$, as well as the length $L$ of DNA chain.
Meanwhile, the dependences of mean FPT on target distance $d_{\rm target}$ and chain distance $d_{\rm chain}$ are also obtained numerically. Here $d_{\rm target/chain}$ is the distance between initial position of protein molecule and the target binding site/DNA chain. In the following subsection, we will first show that there exists a critical value of the one-dimensional diffusion rate $D_1$, with which the mean FPT is almost independent of the binding rate $k_b$ and the unbinding rate $k_d$.

\subsection{The critical value of one-dimensional diffusion rate $D_1$}
\label{subsecThecutoff}
\begin{figure}
  \centering
  \includegraphics[width=10cm]{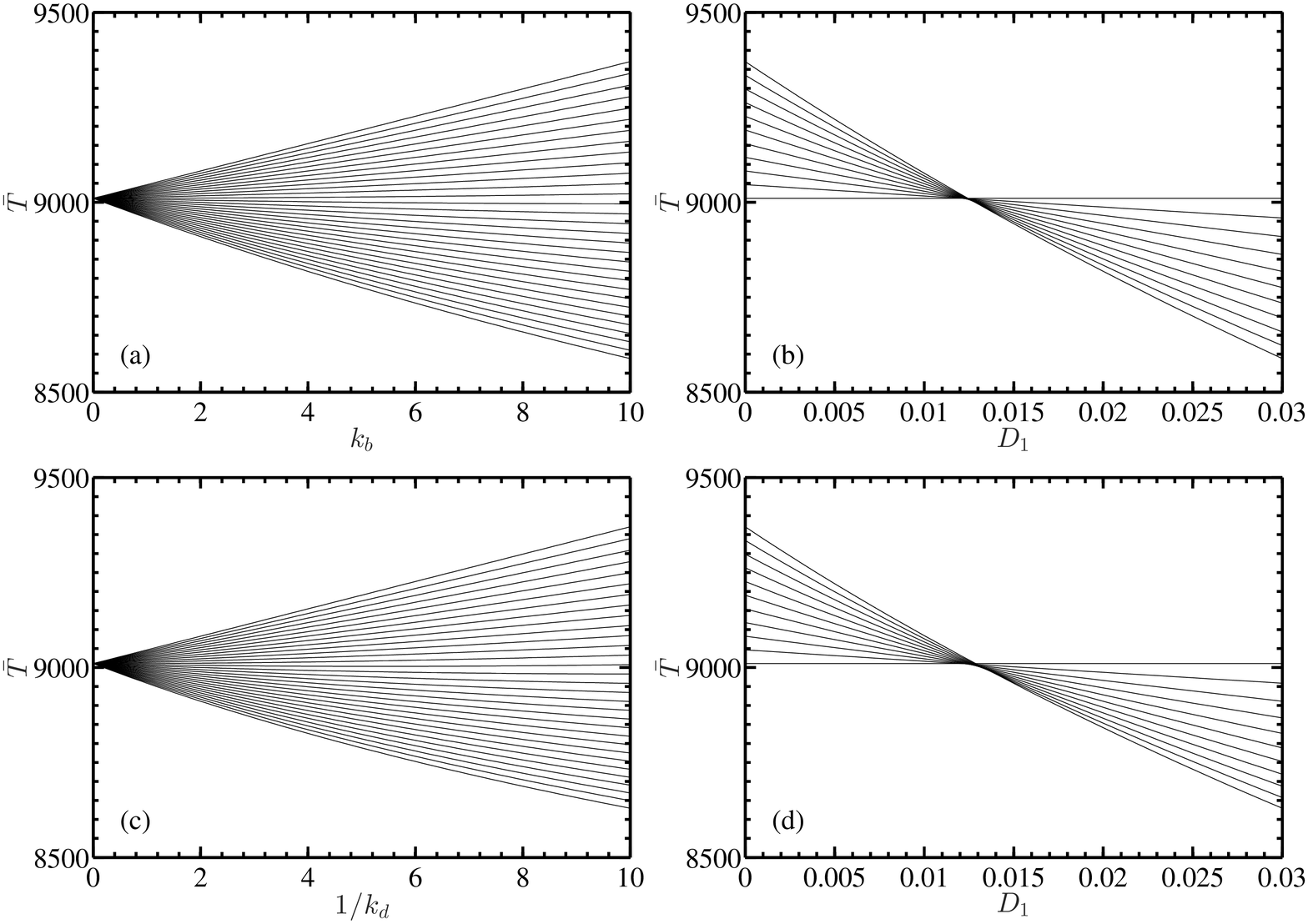}\\
  \caption{The average of mean FPT $\bar{T}$ of a protein in cell space to find its target locus on DNA chain, as functions of binding rate $k_b$ {\bf (a)}, diffusion rate $D_1$ on DNA chain {\bf (b,d)}, and the inverse of detachment rate, $1/k_d$ {\bf (c)}. In {\bf (a,c)}, the values of $D_1$ corresponding to curves from the top down are 0 to 0.03 with increment $0.001$.  In {\bf (b)} the values of $k_b$ corresponding to curves from the bottom up (according to the order at $D_1=0$) are 0 to 10 with increment 0.1. While in {\bf (d)} the values of $1/k_d$ corresponding to curves from the bottom up (according to the order at $D_1=0$) are 0 to 10 with increment 0.1. Other parameter values used in calculations are $M=N=101$, $L=51$, $D_2=1$, $k_t=10$, and $k_d=1$ in {\bf (a,b)}, $k_b=1$ in {\bf (c,d)}. The plots in {\bf (b,d)} show that there exists a critical value of $D_1$, at which the average of mean FPT $\bar{T}$ is insensitive to binding rate $k_b$ and detachment rate $k_d$.
  }\label{FigTMk1k_b_and_TMk1d_date_20160328mean}
\end{figure}

The average of mean FPTs $\bar{T}$ as functions of $D_1$, $k_b$ and $1/k_d$ are plotted in Fig. \ref{FigTMk1k_b_and_TMk1d_date_20160328mean}. Fig. \ref{FigTMk1k_b_and_TMk1d_date_20160328mean}{\bf (a)} shows that, for small values of diffusion rate $D_1$, $\bar{T}$ increases with binding rate $k_b$. While for large values of $D_1$, $\bar{T}$ decreases with $k_b$. Which means that one-dimensional sliding along DNA chain helps to increase the target-search efficiency of a protein only when its sliding speed on DNA chain is large enough. The plots in Fig. \ref{FigTMk1k_b_and_TMk1d_date_20160328mean}{\bf (b)} show that, $\bar{T}$ always decreases with the diffusion rate $D_1$. But there exists a critical value $D_1^*$, for $D_1<D_1^*$ the binding of protein to DNA chain will decrease the search efficiency, and vice versa.
Similar results can be obtained from the plots in Fig. \ref{FigTMk1k_b_and_TMk1d_date_20160328mean}{\bf (c,d)}. Which show that for small values of $D_1$, $\bar{T}$ decreases with unbinding rate $k_d$, while for large values of $D_1$, $\bar{T}$ increase with $k_d$. Therefore, all the plots in  Fig. \ref{FigTMk1k_b_and_TMk1d_date_20160328mean} imply that whether or not one-dimensional sliding along DNA chain is helpful to the target-search process depends on the values of diffusion rate $D_1$. It is helpful only when $D_1$ is larger than the critical value $D_1^*$. For $D_1=D_1^*$, the average of mean FPT $\bar{T}$ is almost insensitive to the change of binding rate $k_b$ and unbinding rate $k_d$, see also plots in Fig. \ref{FigTMk1k2_and_TMk1cl_and_TMk_bd_and_PMk1k2_date_20160328mean}{\bf (c)}. In other words, when $D_1$ takes this critical value, how often and how long a protein slides along the DNA chain have little influence on the time spent by it to reach its target binding site on DNA. Note that we keep the two-dimensional diffusion rate $D_2=1$ in all plots of Fig. \ref{FigTMk1k_b_and_TMk1d_date_20160328mean}. Obviously the value of $D_1^*$ will increase with $D_2$.

To know if the properties obtained for the average of mean FPT $\bar{T}$ from Fig. \ref{FigTMk1k_b_and_TMk1d_date_20160328mean} hold for mean FPT $T_{i,j}$, similar figures for typical mean FPT $T_{i,j}$ are plotted in Figs. \ref{FigTMk1k_b_and_TMk1d_date_20160328p11}-\ref{FigTMk1k_b_and_TMk1d_date_20160328pmm}, and Figs. \ref{FigTMk1k2_and_TMk1cl_and_TMk_bd_and_PMk1k2_date_20160328p11}{\bf (c)}-\ref{FigTMk1k2_and_TMk1cl_and_TMk_bd_and_PMk1k2_date_20160328pmm}{\bf (c)}, with $(i,j)=(1,1), ((M+1)/2,1), (1,(N+1)/2)$, and $((M+1)/2,(N+1)/2)$ respectively. From these plots we can conclude that $T_{i,j}$, the mean FPT of a protein initiated at position $(i,j)$ in the cell space to find its target site on DNA chain, has the same properties as those of $\bar{T}$. The plots in Figs. \ref{FigTMk1k_b_and_TMk1d_date_20160328p11}-\ref{FigTMk1k_b_and_TMk1d_date_20160328pmm} show that, for mean FPT $T_{{(M+1)}/{2}, {(N+1)}/{2}}$, whose initial position is exactly the same as the target site on DNA chain (but lies in cell space, see Fig. \ref{modeldiagram2}), the corresponding critical value of $D_1$ is $D_1^*=0.00752$. While for mean FPTs $T_{1,1}$, $T_{{(M+1)}/{2},1}$, $T_{1,{(N+1)}/{2}}$, the critical value is $D_1^*=0.0151$, which is the same as that obtained from the average of mean FPT $\bar{T}$. This implies a fact that the properties of $\bar{T}$ are mainly determined by the mean FPTs $T_{i,j}$ with initial positions $(i,j)$ near the boundary of the cell space. This fact is due to the effect of high dimension that for a given bounded domain, most of its points are closer to the domain boundary than to the domain center \cite{HastieTibshirani2011}. This high dimension effect appears in other properties of mean FPT as well.
\begin{figure}
  \centering
  \includegraphics[width=10cm]{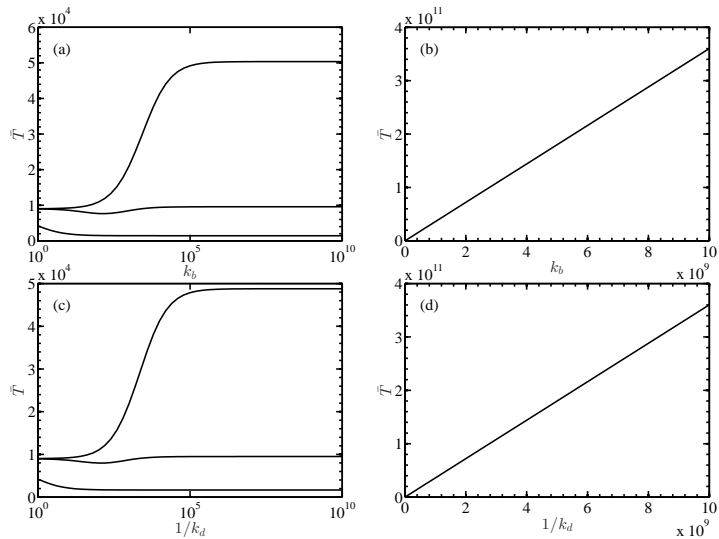}\\
  \caption{Typical examples of $\bar{T}$ as functions of $k_b$ {\bf (a,b)} and $1/k_d$ {\bf (c,d)}, $\large{\textcircled{\small{1}}}$ monotonically increase, $\large{\textcircled{\small{2}}}$ decrease followed by increase, and $\large{\textcircled{\small{3}}}$ monotonically decrease. In {\bf (a,c)}, values of $D_1$ used in calculations are $D_1=0.005,0.03,10$ (from the top down) respectively. {\bf (b,d)} are limit cases with $D_1=0$. Other parameter values used in calculations are $M=N=101$, $L=51$, $D_2=1$, $k_t=10$, and $k_d=1$ in {\bf (a,b)}, $k_b=1$ in {\bf (c,d)}. {\bf (a,c)} show that $\bar{T}$ tends to a constant dependent on $D_1$ as $k_b\to\infty$ or $k_d\to0$.
  }\label{FigTMk1k_bexp_date_20160328}
\end{figure}

Finally, we want to point out that the above results about the critical value of $D_1$ is valid approximately only when the binding rate $k_b$ is not too large and unbinding rate $k_d$ is not too small. In fact, the plots in Fig. \ref{FigTMk1k_bexp_date_20160328}{\bf (a)} show that, if $D_1$ is small/large, then $\bar{T}$ will monotonically increase/decrease with $k_b$, while for intermediate values of $D_1$, $\bar{T}$ first decreases then increases with $k_b$. Nevertheless, as long as $D_1\neq 0$, $\bar{T}$ will always tend to a constant as $k_b\to\infty$. For the special cases with $D_1=0$, $\bar{T}$ increases linearly with $k_b$, see Fig. \ref{FigTMk1k_bexp_date_20160328}{\bf (b)}. For the unbinding rate $k_d$, similar results can be obtained, see Fig. \ref{FigTMk1k_bexp_date_20160328}{\bf (c,d)}.

\subsection{Behaviors of mean FPT within wide range of diffusion rates}
In calculations of previous subsection, to show the existence of critical value of one-dimensional diffusion rate $D_1$, we always fixed the value of two-dimensional diffusion rate $D_2=1$, and varied the one-dimensional diffusion rate $D_1$ in an appropriate range.
In this subsection, we will show how the average of mean FPT $\bar{T}$ changes with $D_1$ and $D_2$ within a large range, i.e., with change from a small value to an extremely large value.
\begin{figure}
  \centering
  \includegraphics[width=10cm]{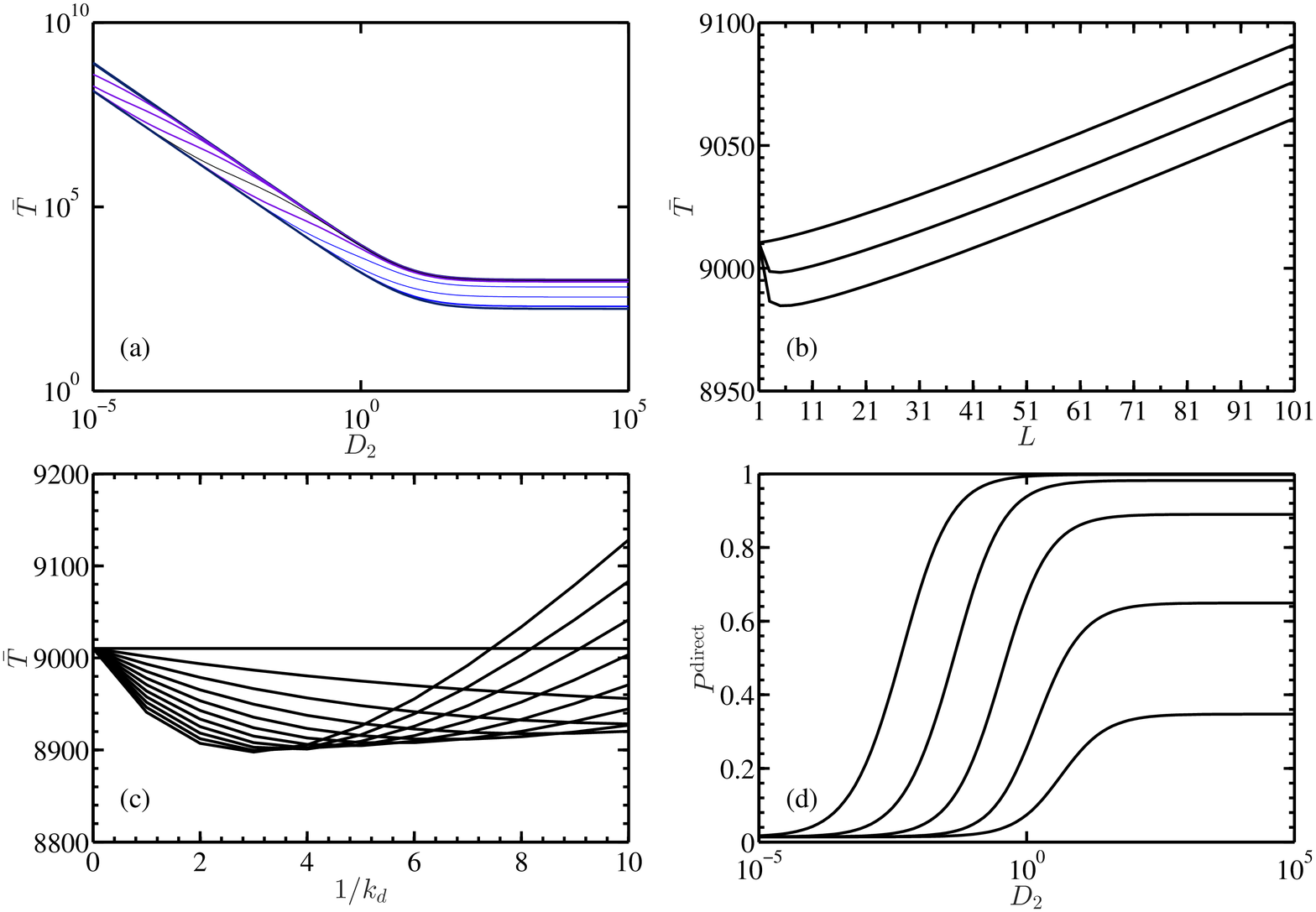}\\
  \caption{{\bf (a)} The value of $\bar{T}$ as a function of $D_2$ with different values of $D_1$. From the top down, the values of $D_1$ used in calculations are $D_1=0,10^{-4},10^{-3},10^{-2},10^{-1},1,10,10^2,10^3,10^{10}$ respectively. Other parameter values used in calculations are $M=N=101$, $L=51$, $k_b=1$, $k_d=1$, and $k_t=10$. {\bf (b)} Typical examples of $\bar{T}$ as functions of length $L$ of DNA chain. Parameter values used in calculations are $M=N=101$, $D_2=1$, $k_b=1$, $k_d=1$, $k_t=10$, and $D_1=0,0.005,0.01$ (from the top down) respectively. For very small $D_1$ (for example $D_1=0$ in the figure), $\bar{T}$ increases monotonically with $L$. While for relatively large $D_1$, $\bar{T}$ decreases rapidly with small values of $L$, and then increases gradually with large $L$. {\bf (c)} $\bar{T}$ as functions of $1/k_d$, with $M=N=101$, $L=51$, $D_2=1$, $k_t=10$, $D_1=0.0151$. The values of $k_b$ corresponding to curves from the bottom up (according to the order near $1/k_d=2$) are 0 to 10 with increment 1. These plots show that $\bar{T}$ is insensitive to $k_b$ and $1/k_d$ in $[0,10]$ when $D_1=0.0151$, see also Figs. \ref{FigTMk1k_b_and_TMk1d_date_20160328mean}{\bf (b,d)}. {\bf (d)} Examples of probability $P^{\rm direct}$ that a protein reaches its target locus on DNA chain through direct binding from cell space but not from the adjacent binding sites of DNA, as functions of two-dimensional diffusion rate $D_2$, with parameter values $M=N=101$, $L=51$, $k_b=1$, $k_d=1$, $k_t=10$, and $D_1=10^{-2},10^{-1},1,10,100$ (from the top down) respectively. These plots show that $P^{\rm direct}$ increases with $D_2$ while decreases with $D_1$.
  }\label{FigTMk1k2_and_TMk1cl_and_TMk_bd_and_PMk1k2_date_20160328mean}
\end{figure}

In Fig. \ref{FigTMk1k2_and_TMk1cl_and_TMk_bd_and_PMk1k2_date_20160328mean}{\bf (a)}, we give a logarithm-logarithm plot of the average mean FPT $\bar{T}$ as a function of $D_2$ with different values of $D_1$. The value of $D_2$ changes in interval $[10^{-5},10^5]$. For the two limit cases, i.e., with $D_1=0$ and $D_1=10^{10}$ (here we use $D_1=10^{10}$ to show the large $D_1$ limit properties), the corresponding curves, denoted by $\bar{T}_{D_1=0}$ and $\bar{T}_{D_1=\infty}$ for convenience, parallel with each other in the logarithm-logarithm scale.
Which means that $\bar{T}_{D_1=0}/\bar{T}_{D_1=\infty}$ is independent of diffusion rate $D_2$. Actually, $\bar{T}_{D_1=0}$ can be regarded as the average of mean FPT $T_{i,j}$ of a protein to reach a target site in cell space, while $\bar{T}_{D_1=\infty}$ can be regarded as the average of $T_{i,j}$ of a protein to reach a DNA chain with $L$ binding sites.

The plots in Fig. \ref{FigTMk1k2_and_TMk1cl_and_TMk_bd_and_PMk1k2_date_20160328mean}{\bf (a)} also show that $\bar{T}$ decreases with two-dimensional diffusion rate $D_2$, and tends to constant as $D_2\to\infty$ while tends to infinity as $D_2\to 0$. As $D_1\to 0$, $\bar{T}$ approaches $\bar{T}_{D_1=0}$ from below, but first for large values of $D_2$ and then for small values of $D_2$. As $D_1\to \infty$, $\bar{T}$ approaches $\bar{T}_{D_1=\infty}$ from above, while first for small values of $D_2$ and then for large values $D_2$.

As functions of rates $D_1$ and $D_2$, mean FPTs $T_{1,1}$, $T_{{(M+1)}/{2},1}$, $T_{1,{(N+1)}/{2}}$ have similar properties as those of $\bar{T}$, see Fig. \ref{FigTMk1k2_and_TMk1cl_and_TMk_bd_and_PMk1k2_date_20160328p11}{\bf (a)}, Fig. \ref{FigTMk1k2_and_TMk1cl_and_TMk_bd_and_PMk1k2_date_20160328pm1}{\bf (a)}, and Fig. \ref{FigTMk1k2_and_TMk1cl_and_TMk_bd_and_PMk1k2_date_20160328p1m}{\bf (a)}.
However, the plots in Fig. \ref{FigTMk1k2_and_TMk1cl_and_TMk_bd_and_PMk1k2_date_20160328pmm}{\bf (a)} show different behavior of mean FPT $T_{{(M+1)}/{2},{(N+1)}/{2}}$. For $D_1=0$, $T_{{(M+1)}/{2},{(N+1)}/{2}}$ is a constant, and is independent of $D_2$. This can be proved theoretically, see Section \ref{appendixIndependencek2} of the Supplementary Material. For $D_1=10^{10}$, $T_{{(M+1)}/{2},{(N+1)}/{2}}$ tends to a constant both as $D_2\to +\infty$ and $D_2\to 0$. For large $D_1$ limit, the protein bound to DNA chain will find its target binding site instantaneously. Therefore, for these cases, $T_{{(M+1)}/{2},{(N+1)}/{2}}$ can be regarded as the FPT of a protein at position $({(M+1)}/{2},{(N+1)}/{2})$ to find the DNA chain. So different from the limit case $D_1=0$, for large $D_1$ limit $T_{{(M+1)}/{2},{(N+1)}/{2}}$ increases with the two-dimensional diffusion rate $D_2$. For $D_2\to 0$, the protein at position $({(M+1)}/{2},{(N+1)}/{2})$ will reach the DNA chain mainly by binding to the target binding site. While for $D_2\to \infty$, the protein will reach each of the site of DNA chain equally.
The curves of $T_{{(M+1)}/{2},{(N+1)}/{2}}$ for intermediate values of $D_1$ approach those of $D_1=0$ and $D_1=10^{10}$ in a same way as previous discussed for $\bar{T}$.

In above discussions, we avoid the special cases with $D_2=0$ since it is a singular point such that $T_{{(M+1)}/{2},{(N+1)}/{2}}=1/k_t$, while the mean FPTs with other initial positions are infinity.

\subsection{The influence of DNA chain length $L$}
The average of mean FPTs $\bar{T}$ as a function of length $L$ of DNA chain is plotted in Fig. \ref{FigTMk1k2_and_TMk1cl_and_TMk_bd_and_PMk1k2_date_20160328mean}{\bf (b)}, with $D_1=0,0.005,0.01$ respectively. There are two cases. For very small values of $D_1$ (see the case with $D_1=0$ in Fig. \ref{FigTMk1k2_and_TMk1cl_and_TMk_bd_and_PMk1k2_date_20160328mean}{\bf (b)}, we claim that this is not the only case but just a typical one with very small $D_1$ value), $\bar{T}$ increases with $L$ monotonically. Which implies that binding to DNA chain always hinders the search of the target site. On the other hand, for relatively large values of $D_1$ (see the cases with $D_1=0.005,0.01$ in Fig. \ref{FigTMk1k2_and_TMk1cl_and_TMk_bd_and_PMk1k2_date_20160328mean}{\bf (b)}), $\bar{T}$ decreases rapidly for small values of $L$, and then increases gradually for large values of $L$. An intuitionistic explanation is that, for small values of $L$, the increase of chain length gives the protein more chances to bind to DNA chain near the target binding site, thereby increases the search efficiency. For large values of $L$, although bindings to DNA chain happen more frequently, most of them are far from the target site, thereby will decrease the search efficiency. One can image that the optimal length $L^*$ of DNA chain, with which the average $\bar{T}$ of mean FPTs reaches its minimum, increases with one-dimensional diffusion constant $D_1$. It has been experimentally found that, by shortening the flanking DNA (the part without target promoter binding site), the rate of promoter binding does not change significantly \cite{Friedman2013PNAS}. This may due to the negative but slight influence of protein bindings to DNA sites far from the target promoter.

Similar results can be obtained for the typical examples of mean FPTs $T_{1,1}$, $T_{{(M+1)}/{2},1}$, $T_{1,{(N+1)}/{2}}$, and $T_{{(M+1)}/{2},{(N+1)}/{2}}$, see Figs. \ref{FigTMk1k2_and_TMk1cl_and_TMk_bd_and_PMk1k2_date_20160328p11}{\bf (b)}, \ref{FigTMk1k2_and_TMk1cl_and_TMk_bd_and_PMk1k2_date_20160328pm1}{\bf (b)}, \ref{FigTMk1k2_and_TMk1cl_and_TMk_bd_and_PMk1k2_date_20160328p1m}{\bf (b)}, and \ref{FigTMk1k2_and_TMk1cl_and_TMk_bd_and_PMk1k2_date_20160328pmm}{\bf (b)} respectively.

\subsection{Mean FPT as functions of distances $d_{\rm target}$ and $d_{\rm chain}$}
\begin{figure}
  \centering
  \includegraphics[width=10cm]{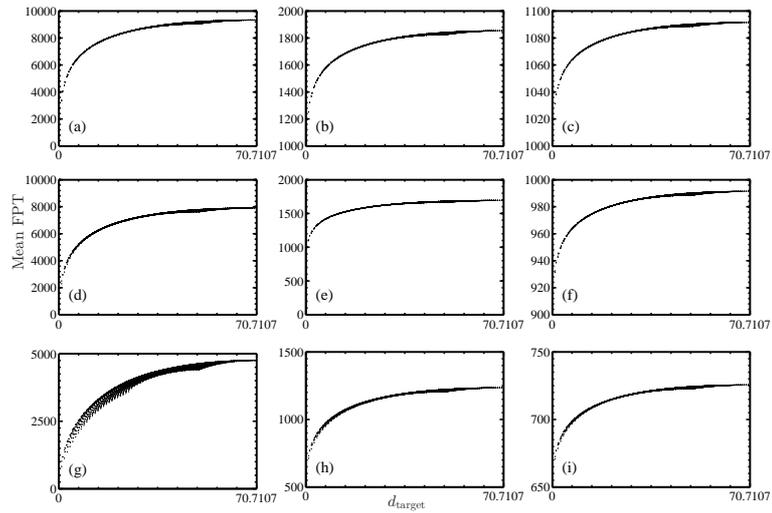}\\
  \caption{The mean FPT $T_{i,j}$ as a function of distance $d_{\rm target}$ between initial position $(i,j)$ and target locus on DNA. The diffusion constant $D_1$ used in calculations of each row of subfigures is $D_1=0.1,1,10$ respectively (from the top down), and $D_2$ used in each column is $D_2=1,10,100$ (from left to right). Except the special cases with high values of $D_1$ and low values of $D_2$, see subfigure {\bf (g)}, the mean FPT increases with $d_{\rm target}$ roughly, and tends to a constant for large values of distance $d_{\rm target}$.
  }\label{FigTMk1k2Dis_date_20160328target}
\end{figure}

Examples of scatter diagram of the mean FPT as a function of $d_{\rm target}$ with different values of $D_1$ and $D_2$ are plotted in Fig. \ref{FigTMk1k2Dis_date_20160328target}. Roughly speaking, the mean FPT $T$ increases with the distant $d_{\rm target}$. Except the special cases where $D_1$ is large while $D_2$ is small, see Fig. \ref{FigTMk1k2Dis_date_20160328target}{\bf (g)}. Since for these cases, the sliding of protein on DNA chain is fast, and therefore the mean FPT is considerably influenced by the binding process of protein to DNA chain. From the plots in Fig. \ref{FigTMk1k2Dis_date_20160328target}, one can find that although the value of mean FPT $T$ changes with diffusion rates $D_1$ and $D_2$, the shape of function $T(d_{\rm target})$ does not change significantly.

\begin{figure}
  \centering
  \includegraphics[width=10cm]{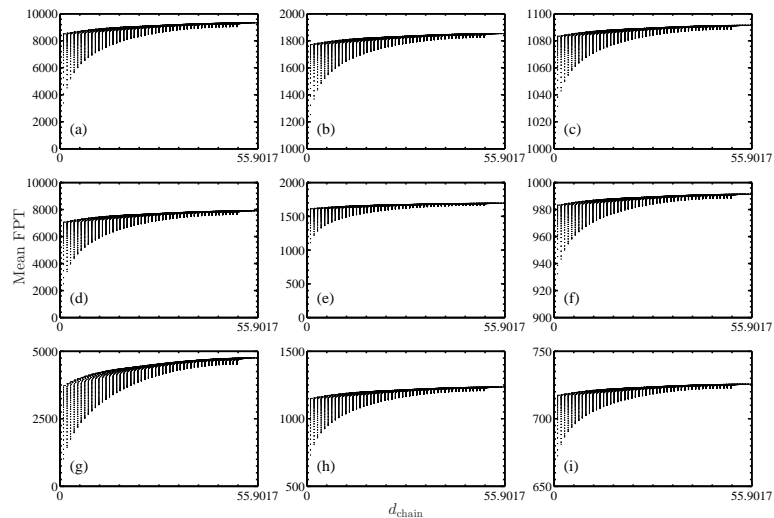}\\
  \caption{The mean FPT $T_{i,j}$ as a function of distance $d_{\rm chain}$ between initial position $(i,j)$ and the DNA chain. The diffusion constant $D_1$ used in calculations of each row is $D_1=0.1,1,10$ (from the top down), and $D_2$ used in each column is $D_2=1,10,100$ (from left to right). }\label{FigTMk1k2Dis_date_20160328chain}
\end{figure}

One can image that if the one-dimensional sliding on DNA chain can decrease the search time remarkably, then the mean FPT $T$ of a protein will strongly depend on the distance $d_{\rm chain}$ between the initial position of protein and the DNA chain. Here $d_{\rm chain}$ is defined as the minimum of the distances between the initial position of protein and all binding sites of the DNA chain. The plots in Fig. \ref{FigTMk1k2Dis_date_20160328chain} show that the mean FPT $T$ does not increase monotonically with the distance $d_{\rm chain}$. However, one can find that the maximal value $T_{\max}$ and minimal value $T_{\min}$, as well as the average value $T_{\rm average}$ of mean FPT $T$ do increase with distance $d_{\rm chain}$. Where $T_{\max}(d_{\rm chain}):=\max\{T_{i,j}|\textrm{the distance between position $(i,j)$ and DNA chain is } d_{\rm chain}\}$, and $T_{\min}$ and $T_{\rm average}$ are defined similarly.

From all plots in Figs. \ref{FigTMk1k2Dis_date_20160328target} and \ref{FigTMk1k2Dis_date_20160328chain}, as well as further detailed analyses of the calculation results, we conclude that the mean FPT $T(d_{\rm target}, d_{\rm chain})$ increases with both distance $d_{\rm target}$ and distance $d_{\rm chain}$ roughly. For large values of $D_2$ while small values of $D_1$, the influence of distance $d_{\rm chain}$ is negligible, which implies that the one-dimensional sliding on DNA chain has almost no contribution to increase the search efficiency of protein.

\section{The direct binding probability $P^{\rm direct}$ }\label{diffusiondependence}
In this section, we will discuss the dependence of direct binding probability $P^{\rm direct}$ on diffusion rates $D_1$ and $D_2$, see Section \ref{SecAmethod} of the Supplementary Material for the method used in this study to get $P^{\rm direct}$. As has been defined before, $P^{\rm direct}$ describes the probability that protein reaches its target binding site through its adjacent position in cell space but not its nearest neighbor sites on DNA chain. For convenience, we define $P_{i,j}^{\rm direct}$ as the direct binding probability of a protein initiated at position $(i,j)$ in the cell space, and then $P^{\rm direct}$ is obtained as the average of $P_{i,j}^{\rm direct}$,
\begin{eqnarray}
P^{\rm direct}=\frac{\sum_{i=1}^M\sum_{j=1}^N P_{i,j}^{\rm direct}}{MN}.
\end{eqnarray}

For given values of $L$, $k_b$, $k_d$ and $k_t$, the numerical results of $P^{\rm direct}$ are plotted in Fig. \ref{FigTMk1k2_and_TMk1cl_and_TMk_bd_and_PMk1k2_date_20160328mean}{\bf (d)} for different values of diffusion rates $D_1$ and $D_2$. We found that $P^{\rm direct}$ increases with two-dimensional diffusion rate $D_2$ while decreases with one-dimensional diffusion rate $D_1$. Which implies that for large $D_2$ but small $D_1$, protein will reach its target binding locus mainly through two-dimensional diffusion, and the contribution of one-dimensional sliding along DNA chain is not significant. One the contrary, for small $D_2$ but large $D_1$, one-dimensional sliding along DNA plays main role on the target search process of protein. One can also find from Fig. \ref{FigTMk1k2_and_TMk1cl_and_TMk_bd_and_PMk1k2_date_20160328mean}{\bf (d)} that the limit value of $P^{\rm direct}$ for $D_2\to \infty$ depends on $D_1$, while the limit value for $D_2\to 0$ does not. The properties of typical examples $P_{1,1}^{\rm direct}$, $P_{(M+1)/2,1}^{\rm direct}$ and $P_{1,(N+1)/2}^{\rm direct}$ are similar as those of $P^{\rm direct}$, see Figs. \ref{FigTMk1k2_and_TMk1cl_and_TMk_bd_and_PMk1k2_date_20160328p11}{\bf (d)}, \ref{FigTMk1k2_and_TMk1cl_and_TMk_bd_and_PMk1k2_date_20160328pm1}{\bf (d)}, and \ref{FigTMk1k2_and_TMk1cl_and_TMk_bd_and_PMk1k2_date_20160328p1m}{\bf (d)}. Nevertheless, the plots in Fig. \ref{FigTMk1k2_and_TMk1cl_and_TMk_bd_and_PMk1k2_date_20160328pmm}{\bf (d)} show that the behavior of $P_{(M+1)/2, (N+1)/2}^{\rm direct}$ is different. It increases to value one monotonically as two-dimensional diffusion rate $D_2$ decreases, but still decreases with one-dimensional diffusion rate $D_1$. The behavior of average value $P^{\rm direct}$ is similar as those of $P_{i,j}^{\rm direct}$ with $(i,j)$ near the boundary of the two-dimensional cell space. As mentioned  in Section \ref{subsecThecutoff}, this is due to the effect of high dimension that most of the cell positions are closer to the cell boundary than to the cell center, which is assumed to be the position of the target binding locus of DNA chain in this study.

\section{Concluding and remarks}
In this study, a simple chain-space coupled model is employed to discuss the search process of a protein molecule in cell space to its target locus on DNA chain. The protein molecule may be RNA polymerase or transcription factor, and the binding of it to certain locus of DNA is essential to regulate the expression of gene. How these protein molecules find their corresponding target loci on DNA remains unclear. In our study, the DNA chain is simplified to be a one-dimensional lattice, and the cell space is simplified to be a rectangular lattice. The mean first-passage time (FPT) is chosen as one criterion to evaluate the search efficiency.

Our results show that there exists one critical value of the one-dimensional diffusion rate $D_1$. With which the search efficiency of a protein molecule is almost independent of the frequency and time that the protein molecule slides along DNA chain. If the value of $D_1$ is larger than this critical value, the one-dimensional sliding along DNA chain will be helpful to increase the search efficiency of protein, while for $D_1$ lower than this critical value, sliding along DNA chain will have no contribution to the search process.

Meanwhile, this study found that the search efficiency of protein increases first and then decreases with the length of DNA chain. Which implies that only bindings to sites of DNA chain near the target locus can help to increase the search efficiency. This is consistent with the claim given in \cite{Friedman2013PNAS} that most proteins bind the target locus of DNA chain without long-range one-dimensional sliding.

The probability $P^{\rm direct}$ that a protein molecule reaches its target locus through direct binding from nearby position in cell space (but not adjacent sites on DNA chain) is also discussed in this study. Our results show that $P^{\rm direct}$ increases with two-dimensional diffusion rate $D_2$ while decreases with one-dimensional diffusion rate $D_1$. Therefore, for high values of $D_2$ but low values of $D_1$, protein molecule will reach its target locus mainly through diffusion in cell space. One the contrary, for low values of $D_2$ but high values of $D_1$, protein molecule will reach its target mainly through one-dimensional sliding along DNA chain.

The results obtained in this study will help to understand the target search process in cells during gene expression. Which show that how RNA polymerase or transcription factor reaches their binding sites on DNA, i.e. mainly through high dimensional diffusion or with the help of one-dimensional sliding along DNA, depends on the detailed environment of cells (such as diffusion constants) as well as the properties of RNA polymerase or transcription factor (binding/unbinding rate to/from DNA, and {\it etc}).

\section*{Author Contributions}
Y.Z. designed the research; J.L. and Y.Z. performed the research and wrote the paper, J.L. wrote programs.

\section*{Acknowledgments}
This study was supported by the Natural Science Foundation of China (Grant No. 11271083).


\end{document}